\documentstyle[namedreferences,NATO,epsf]{CRCKAPB}

\begin{opening}
\title{Magnetic Field Evolution in Accreting White Dwarfs}
\author{Andrew Cumming}
\institute{Hubble Fellow, Department of Astronomy and Astrophysics,\\University of California, Santa Cruz, CA 95064, USA}
\end{opening}

\begin{document}

\begin{abstract}
I discuss the evolution of the magnetic field of an accreting white
dwarf. I show that the ohmic decay time is $7$--$12$ billion years for
the lowest order decay mode, almost independent of core temperature or
mass. I then show that the magnetic field structure is substantially
altered by accretion if the white dwarf mass increases at a rate
$>\dot M_c\approx (1$--$5)\times 10^{-10}\ M_\odot\ {\rm yr^{-1}}$. I
discuss the implications of this result for observed systems,
including the possible evolutionary link between AM Hers and
intermediate polars.
\end{abstract}

\section{Introduction}

Calculations of ohmic decay in an isolated, cooling white dwarf show
that the magnetic field changes little over its lifetime (Chanmugam \&
Gabriel 1972; Fontaine, Thomas, \& van Horn 1973; Wendell, van Horn,
\& Sargent 1987). However, an accreting white dwarf may substantially
increase its mass on a timescale much shorter than the ohmic decay
time, raising the question of what happens to its magnetic
field. Here, I describe calculations which show that the field
structure may be substantially affected by accretion if the accretion
is rapid enough, $\dot M>\dot M_c\approx (1$--$5)\times 10^{-10}\
M_\odot\ {\rm yr^{-1}}$. At lower accretion rates, ohmic diffusion
allows the field to ``keep up'' with the accretion flow. This critical
accretion rate lies in the middle of observed rates, with interesting
implications for observations. A detailed account of this work may be
found in Cumming (2002).

A major uncertainty in our understanding of accreting white dwarfs is
whether the white dwarf mass increases or decreases with time. The
answer depends on the amount of mass ejected by classical novae, an
uncertain quantity both observationally and theoretically. In the
calculations presented here, I assume the white dwarf mass increases
with time.

\section{Ohmic Decay in Liquid White Dwarfs}

Compressional heating maintains the core of most accreting white
dwarfs in a liquid state, with core temperatures $>10^7\ {\rm K}$
(Nomoto 1982). In this case, the ohmic decay timescale is almost
independent of white dwarf mass and core temperature.

The timescale for ohmic decay is $t_{\rm ohm}=4\pi\sigma L^2/c^2$,
where $\sigma$ is the electrical conductivity, and $L$ the lengthscale
over which the magnetic field changes. The larger the conductivity,
the less dissipation, giving a longer ohmic time. The insensitivity to
core temperature comes about because in a liquid, the electrical
conductivity is set by electron-ion scattering, which depends mainly
on the electron Fermi energy (and therefore density) rather than
temperature, $\sigma\propto\rho^{1/3}$ at the center. In contrast, a
cooling white dwarf with a solid core has conductivity set by
electron-phonon collisions, and the phonon density depends sensitively
on temperature. To get the rough scaling with mass, assume a
mass-radius relation $R\propto M^{-1/3}$, giving a mean density
$\rho\propto M^2$, and therefore $\sigma\propto M^{2/3}$. The scaling
with mass then drops out of the ohmic time, $t_{\rm ohm}\propto \sigma
R^2=$ constant.

Following the calculations of Wendell et al.~(1987) for isolated white
dwarfs, I have calculated the ohmic decay modes for detailed models of
accreting white dwarfs. This calculation takes into account the
variations of $L$ and $\sigma$ at different radii. The lowest order
ohmic decay mode has a decay time that varies by a factor of 2 over
mass, $\tau=7$--$12$ billion years.

\section{Calculation of the Critical Accretion Rate}

\begin{figure}
\begin{centering}
\epsfxsize=3.5in \epsfbox{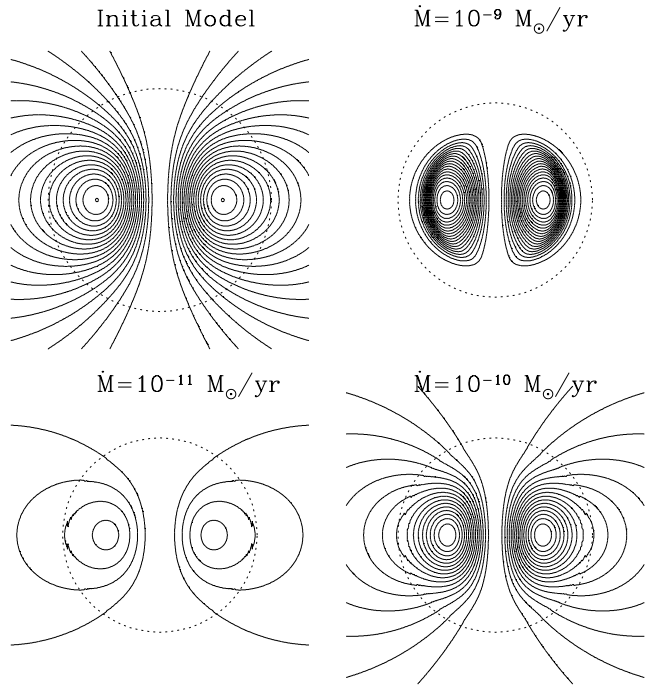}
\caption{Magnetic field lines before and after accretion of $0.1\
M_\odot$ onto a $0.6\ M_\odot$ white dwarf at three different rates,
$\dot M=10^{-11}, 10^{-10},$ and $10^{-9}\ M_\odot\ {\rm
yr^{-1}}$. For $\dot M\leq 10^{-10}\ M_\odot\ {\rm yr^{-1}}$, the
number density of field lines decreases due to ohmic decay. For $\dot
M=10^{-9}\ M_\odot\ {\rm yr^{-1}}$, magnetic flux is advected into the
interior. Field lines with the same interior flux are shown in each
plot.}
\end{centering}
\end{figure}

I adopt two approaches to calculate the critical accretion rate $\dot
M_c$ above which ohmic diffusion can no longer keep up with accretion,
and we expect accretion to change the white dwarf field structure. The
first, following work on accreting neutron stars (Cumming, Zweibel, \&
Bildsten 2001), is to compare local timescales for accretion and ohmic
diffusion as a function of depth in the white dwarf envelope. I find
that the ratio $t_{\rm diff}/t_{\rm accr}$ is insensitive to depth,
and is unity for $\dot M=(1$--$5)\times 10^{-10}\ M_\odot\ {\rm
yr^{-1}}$.

The second approach is to calculate the global time-dependent
evolution of the field under the joint action of accretion and ohmic
diffusion. I consider an axisymmetric dipole magnetic field with
spherical accretion. The evolution of the field is described by a 1D
advection-diffusion equation which is straightforward to evolve in
time. Figure 2 shows the initial and final field configurations after
accreting $0.1\ M_\odot$ onto a $0.6\ M_\odot$ white dwarf at three
different rates. For $\dot M\leq 10^{-10}\ M_\odot\ {\rm yr^{-1}}$,
the field structure is unaffected by accretion, but the number density
of field lines decreases due to ohmic decay. For $\dot M=10^{-9}\
M_\odot\ {\rm yr^{-1}}$, however, the field structure is substantially
changed as magnetic flux is advected inwards by accretion.

\section{Implications for Observed Systems}

I have shown that the surface magnetic field is reduced by accretion
if the white dwarf mass increases at a rate $\dot M>(1$--$5)\times
10^{-10}\ M_\odot\ {\rm yr^{-1}}$. This has a number of implications
for observed systems.

{\it An evolutionary connection between AM Hers and intermediate
polars?}  The $\dot M$ estimates of Warner (1995) show that AM Hers
accrete at low rates, $\dot M\approx 5\times 10^{-11}\ M_\odot\ {\rm
yr^{-1}}$, whereas the IPs accrete more rapidly, $\dot M\approx
(0.2$--$4)\times 10^{-9}\ M_\odot\ {\rm yr^{-1}}$. Interestingly, the
critical rate $\dot M_c$ lies between these two values, suggesting the
possibility that the magnetic fields in IPs have been reduced by
accretion. This would allow for an evolutionary connection between IPs
and AM Hers, which is suggested by their different orbital period
distributions, but has difficulties if $B$ is constant (Hameury et
al.~1987; King \& Lasota 1991), since IPs are believed to have weaker
fields than AM Hers. A drop in $\dot M$, e.g.~at $P_{\rm orb}\approx
3$--$4 {\rm h}$ would allow the magnetic field to reemerge, on a
timescale $\approx 3\times 10^8\ {\rm yr}\ (\Delta M/0.1\
M_\odot)^{7/5}$ ($\Delta M$ is the amount of accreted mass), similar
to the time for a non-magnetic CV to cross the period gap. The picture
is presumably more complex, for example within AM Hers, no correlation
is seen between $B$ and orbital period (Wickramasinghe \& Ferrario
2000). However, the fact that the magnetic field may change with time
should be kept in mind when considering the evolution of these
systems.

{\it Rapidly accreting systems (``Type Ia progenitors'')} The
supersoft X-ray sources and symbiotic systems have $\dot M\gg\dot
M_c$, and are believed to be stably burning the accreted matter,
increasing the white dwarf mass. We know little about the magnetic
fields of these white dwarfs. Pulsations are seen in the symbiotic Z
And (Sokoloski \& Bildsten 1999), and a supersoft source in M31
(Osborne et al.~2001; King, Osborne, \& Schenker 2002). If due to
magnetic accretion, then $B\approx 10^7\ {\rm G}$ in each case. We
expect accretion to have a substantial effect on the magnetic field,
so these are interesting systems to consider further.

{\it Where are the $10^9\ {\rm G}$ accreting white dwarfs?} Whereas
isolated white dwarf fields extend up to $10^9\ {\rm G}$, most AM Hers
have $B<10^8\ {\rm G}$, with the highest being $2.8\times 10^8\ {\rm
G}$ (AM Uma). This may be due to selection effects (Wickramasinghe \&
Ferrario 2000). Some difference is to be expected from the different
ohmic decay properties --- accreting white dwarfs are liquid and have
shorter decay times. This could account for a factor of 2 difference
over a Hubble time.

I thank D. Wickramasinghe, H. van Horn, and E. Zweibel for useful
discussions. This work was supported by NASA Hubble Fellowship grant
HF-01138 awarded by the Space Telescope Science Institute, which is
operated by the Association of Universities for Research in Astronomy,
Inc., for NASA, under contract NAS 5-26555.

\end{document}